\documentclass[12pt]{article}
\usepackage[T1]{fontenc}
\usepackage[latin1]{inputenc}

\usepackage{epsfig}

\newcommand{\lsim}{\,{\buildrel < \over {_\sim}}\,}
\newcommand{\gsim}{\,{\buildrel > \over {_\sim}}\,}

\newcommand{\shat}{{\hat s}}
\newcommand{\that}{{\hat t}}

\begin{document}

\begin{titlepage}
\begin{flushright}
JYFL-6/01\\
hep-ph/0104124\\
April 2001
\end{flushright}
\vfill

\begin{centering}

{\bf OBTAINING THE NUCLEAR GLUON DISTRIBUTION FROM HEAVY QUARK DECAYS TO 
LEPTON PAIRS IN p$A$ COLLISIONS}\\

\vspace{0.5cm}
K.J. Eskola$^{\rm a,b}$\footnote{kari.eskola@phys.jyu.fi},
V.J. Kolhinen$^{\rm a}$\footnote{vesa.kolhinen@phys.jyu.fi} and
R. Vogt$^{\rm c}$\footnote{vogt@lbl.gov}

\vspace{1cm}
{\em $^{\rm a}$Department of Physics, University of Jyv\"askyl\"a,\\
P.O. Box 35, FIN-40351 Jyv\"askyl\"a, Finland}
\vspace{0.3cm}

{\em $^{\rm b}$Helsinki Institute of Physics,\\
P.O. Box 64, FIN-00014 University of Helsinki, Finland}
\vspace{0.3cm}

{\em $^{\rm c}$Lawrence Berkeley National Laboratory, Berkeley, CA 94720, 
USA,\\  and \\ Physics Department, University of California, Davis, CA 95616,
USA}

\vspace{0.3cm}

\vspace{1cm} 
{\bf Abstract} \\ 
\end{centering}

We have studied how lepton pairs from decays of heavy-flavoured mesons
produced in p$A$ collisions can be used to determine the modifications
of the gluon distribution in the nucleus.  Since heavy quark
production is dominated by the $gg$ channel, the ratio of correlated
lepton pair cross sections from $D\overline D$ and $B\overline B$
decays in p$A$ and pp collisions directly reflects the ratio $R_g^A
\equiv f_g^A/f_g^p$.  We have numerically calculated the lepton pair
cross sections from these decays in pp and p$A$ collisions at SPS,
RHIC and LHC energies. We find that ratio of the p$A$ to pp cross
sections agrees quite well with the input $R_g^A.$ Thus, sufficiently
accurate measurements could be used to determine the nuclear
modification of the gluon distribution over a greater range of $x$ and
$Q^2$ than presently available, putting strong constraints on models.

\vspace{0.3cm}\noindent

\vfill
\end{titlepage}

%
%

\section{Introduction}

Inclusive differential cross sections of hard scatterings in
proton-nucleus collisions at high energies are, to first
approximation, computable assuming factorization. For processes not
sensitive to isospin effects, such as heavy quark production, these
cross sections can be expressed as
\begin{eqnarray}
& \displaystyle 
   d\sigma(Q^2,\sqrt s)_{{\rm p}A\rightarrow a+X} =
   \sum_{i,j=q,\overline q,g} 
   f_i^{\rm p}(x_1,Q^2)\otimes Af_j^A(x_2,Q^2)
   \otimes d\hat \sigma(Q^2,x_1,x_2)_{ij\rightarrow a+x}
\label{hardpA}
\end{eqnarray}
where $\hat \sigma(Q^2,x_1,x_2)_{ij\rightarrow a+x}$ are the
perturbatively calculable partonic cross sections for producing parton
$a$ at scale $Q^2\gg\Lambda^2_{\rm QCD}$, $x_1$ and $x_2$ are the
momentum fractions of the partons involved in the hard scattering,
while $f_i^{\rm p}$ and $f_i^A$ are the distributions of parton $i$ in
a free proton and a nucleus with mass number $A$ respectively.  In the
high-$Q^2$ limit, possible multiple scatterings of individual partons
\cite{QIU,SCHAFER} are neglected but the parton distributions will be
modified in the nucleus \cite{EKR98,EKS98}. Equation (\ref{hardpA})
may be generalized to nucleus-nucleus collisions but we focus only on
$pA$ collisions in this paper.

The measurements of the structure function $F_2^A$ in deep inelastic
lepton-nucleus scattering (DIS) clearly show that, compared to
deuterium, D, the ratios $R_{F_2}^A \equiv F_2^A/F_2^{\rm D}$, differ
from unity.  As nuclear effects in D are negligible, this indicates
that the parton distributions of bound nucleons are different from
those of a free proton, $R_i^A(x,Q^2)\equiv f_i^A(x,Q^2)/f_i^{\rm
p}(x,Q^2) \neq 1$. The nuclear effects observed in $F_2^A$ are usually
categorized according to the different regions of Bjorken-$x$ as:
shadowing, $R_{F_2}^A \le 1$, at $x \lsim 0.1$; antishadowing,
$R_{F_2}^A \ge 1$, at $0.1 \lsim x \lsim 0.3$; the EMC effect,
$R_{F_2}^A \le 1$, at $0.3 \lsim x\lsim0.7$; and Fermi motion,
$R_{F_2}^A \ge 1$, at $x\rightarrow1$ and beyond. For a review of the
measurements and some models of nuclear effects, see
Ref.~\cite{ARNEODO94}.

The scale evolution of the nuclear parton distributions is, to a first
approximation, similar to that of the free proton.  Thus at scales
$Q^2> Q_0^2\gg\Lambda_{\rm QCD}^2$ the evolution of $f_i^A(x,Q^2)$ can
be described by the Dokshitzer-Gribov-Lipatov-Altarelli-Parisi (DGLAP)
evolution equations \cite{DGLAP} if the initial conditions are given
at a fixed initial scale $Q_0^2$.  Such a DGLAP analysis of the
nuclear parton densities \cite{EKR98,EKS98,KJE93} assumes that the
nuclear effects factorize from the parton densities for $Q_0^2\gsim
1$~GeV$^2$ and the subsequent $Q^2$-evolution of the modifications
follows the DGLAP equations, neglecting $\sim 1/Q^2$ corrections
\cite{GLR,MQiu}. The nuclear modifications at $Q^2_0$,
$R_i^A(x,Q_0^2)$, can then be determined based on constraints provided
by the measurements of $F_2^A$ in DIS, the Drell-Yan cross sections in
$pA$ collisions \cite{E772-DY,E866} and conservation of momentum and
baryon number. This procedure was carried out in Ref.~\cite{EKR98}. In
the present work, we utilize the EKS98 parametrization of the nuclear
effects $R_i^A(x,Q^2)$, obtained from Ref.~\cite{EKR98}, released for
public use in Ref.~\cite{EKS98}.

%
%
\begin{figure}[!t]
\centering\includegraphics[width=10cm]{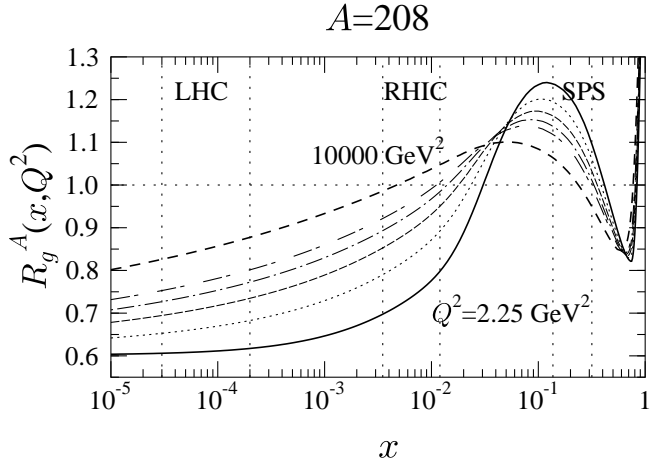}
\caption[a] { \small Scale evolution of the ratio $R_g^A(x,Q^2)$ for an
isoscalar nucleus $A$=208 according to EKS98 \cite{EKR98,EKS98}. The 
ratios are shown as functions of $x$ at fixed values of $Q^2$ 
equidistant in $\log Q^2$: 
2.25~GeV$^2$ (solid),
5.39~~GeV$^2$ (dotted),
14.7~~GeV$^2$ (dashed),
39.9~~GeV$^2$ (dotted-dashed),
108~GeV$^2$ (double-dashed), and 
10000~GeV$^2$ (dashed).
The regions between the vertical dashed lines show the dominant values
of $x_2$ probed by muon pair production from $D\overline D$ at SPS, RHIC 
and LHC energies.
}
\label{RGDIST}
\end{figure}

There remain uncertainties, especially in the determination
of the nuclear gluon distributions. Figure \ref{RGDIST} shows the
gluon modifications, $R_g^A(x,Q^2)$, for a heavy nucleus, $A=208$, as
obtained from a DGLAP analysis \cite{EKR98,EKS98}.  So far there are no
direct constraints on the nuclear gluon distributions. These can,
however, be constrained indirectly through the scale evolution of
$F_2^A(x,Q^2)$ \cite{EKR98,GP96}.  The only measured $Q^2$ dependence
is of the ratio $F_2^{\rm Sn}/F_2^{\rm C}$ \cite{NMC96} at
$0.02<x<0.2$.  At other values of $x$, only the requirement of
stability of the evolution can be used in the perturbative
analysis. Finally, a global limit on the nuclear gluon distribution is
provided by momentum conservation which effectively determines the
level of antishadowing in $R_g^A(x,Q_0^2)$.  More measurement-based
constraints are thus vital for reducing the uncertainties in the
nuclear gluon distributions.  Heavy quark production is one such
process.

The dominant subprocess of heavy quark production in hadronic or
nuclear collisions is $gg \rightarrow Q \overline Q$.  The produced
heavy quarks fragment into mesons which may decay
semi-leptonically. Lepton pairs formed from these decay leptons thus
carry direct information about the input nuclear gluon distribution.
Previous studies can be found in
Refs. \cite{GGRV96-CHARM,LINGYULASSY96}.  At that time, however,
$R_g^A$ was not yet as well constrained as in the EKS98
parameterization.

There are clear advantages for studying the nuclear gluon distribution
in p$A$ interactions.  In p$A$ studies, any possible complications of
dense collective matter produced in ultrarelativistic $AA$ collisions
are avoided. Also, the number of uncorrelated pairs where e.g.\ the
$D$ and $\overline D$ originate in two independent partonic
subcollisions remains rather small.

The purpose of this paper is to study to what extent lepton pair
production from heavy flavor meson decays reflects nuclear
effects on the gluon distributions in p$A$ collisions, especially in
the ratio $\sigma^{{\rm p}A}/\sigma^{{\rm pp}}$.  As the input nuclear gluon
modification, we will use the EKS98 gluon parameterization shown in
Fig.~\ref{RGDIST}. Both correlated lepton pairs, where the $Q\overline
Q$ pair is created in one subcollision, and uncorrelated pairs are
considered.  To probe the ratio $R_g^A(x,Q^2)$ over as wide an $x$
range as possible, we study p$A$ and pp collisions at the SPS ($\sqrt
s= 17.3$ GeV), RHIC ($\sqrt s=200$ GeV) and LHC ($\sqrt s=5500$ GeV)
assuming the same energy as in $AA$ collisions for both pp and p$A$ studies.  
Lepton pair
production from $D\overline D$ decays at the SPS (NA60) will probe the
antishadowing region while at RHIC (PHENIX, STAR) and LHC (ALICE, CMS)
the cross sections are predominantly sensitive to the gluon properties
in the shadowing region, as indicated in Fig.~\ref{RGDIST}.

It should be noted that the present work focuses only on the best-case
scenario in which we assume that the leptons from heavy
meson decays can be identified through vertex displacement. We make no
attempt to simulate the background processes in detail here. This
study, together with Refs.~\cite{GGRV96-CHARM,LINGYULASSY96}, should
provide further motivation for performing more detailed event
simulations in real detectors. Our hope is that the lepton pair cross
sections measured in the future could be used to constrain the
remaining uncertainties in the nuclear gluon distribution in a
model-independent way.

%
%

\section{Formalism}

\subsection{Lepton pair production in p$A$ collisions}

We first present the expression for the differential cross
section of lepton pair production in p$A$ collisions.  To simplify
notation, we refer to generic heavy quarks, $Q$, and heavy-flavoured mesons,
$H$.  In general, the lepton pair production cross section is
\begin{eqnarray}
  \displaystyle\frac{d\sigma^{{\rm p}A \rightarrow l \overline l +X}}
  {dM_{l \overline l} dy_{l \overline l}} & = &
   \displaystyle \int d^3\vec p_{l} d^3\vec p_{\overline l} 
                 \int d^3\vec p_{H} d^3\vec p_{\overline H}
    \,\delta(M_{l \overline l}-M(p_l,p_{\overline l})) 
    \,\delta(y_{l \overline l}-y(p_l,p_{\overline l}))  \nonumber \\
  & & \displaystyle \times
    \frac{d\Gamma^{H \rightarrow l+X}(\vec p_H)}{d^3 \vec p_l}
    \,\, 
    \frac{d\Gamma^{\overline H \rightarrow \overline l+X}(\vec p_{\overline 
     H})}
         {d^3 \vec p_{\overline l}}
    \,\, 
    \frac{d\sigma^{{\rm p}A \rightarrow H \overline H+X}}
         {d^3 \vec p_{H} d^3 \vec p_{\overline H }} \qquad \qquad
    \nonumber \\
  & & \times 
    \theta(y_{\rm min}<y_{l},y_{\overline l}<y_{\rm max})
    \theta(\phi_{\rm min}<\phi_{l},\phi_{\overline l}<\phi_{\rm max}) \, \, . 
   \qquad
    \label{dspAll}
\end{eqnarray}
where $M_{l \overline l}$ 
and $y_{l \overline l}$ are the mass and the rapidity of the
lepton pair, defined as
\begin{eqnarray}
 M(p_l,p_{\overline l}) & = &
    \sqrt{(p_l+p_{\overline l})^2} \, \, , \\
 y(p_l,p_{\overline l}) & = & \displaystyle\frac{1}{2} 
    \ln \left( \frac{(E_l+E_{\overline l})+(p_{lz}+p_{{\overline l}z})}
              {(E_l+E_{\overline l})-(p_{lz}+p_{{\overline l}z})} \right) \,
              \, .
\end{eqnarray}
The decay rate, $d\Gamma^{H \rightarrow l+X}(\vec p_H)/d^3 \vec p_l$, is 
the probability that meson $H$ with momentum $\vec p_H$ decays
to a lepton $l$ with momentum $\vec p_l$.  The $\theta$ functions define
single lepton rapidity and azimuthal angle cuts used to simulate
detector acceptances.

Using a fragmentation function $D^{H}_{Q}$ to describe quark
fragmentation to mesons, the $H \overline H$ production cross section 
can be written as
\begin{eqnarray}
  \frac{d\sigma^{{\rm p}A \rightarrow H \overline H+X}}
       {d^{3} \vec p_H d^3 \vec p_{\overline H}}  &= &
  \displaystyle 
    \int \frac{d^3\vec p_{Q}}{E_{Q}} 
         \frac{d^3\vec p_{\overline Q}}{E_{\overline Q}}
   \, E_{Q}E_{\overline Q}\frac{d\sigma^{{\rm p}A \rightarrow Q 
    \overline Q+X}}
        {d^{3} \vec p_Q d^3 \vec p_{\overline Q}}  
    \int_{0}^{1} dz_1 D^{H}_{Q}(z_1)
    \int_{0}^{1} dz_2 D^{\overline H}_{\overline Q}(z_2) \nonumber \\
   & \times & \, \delta^{(3)}(\vec p_H - z_1 \vec p_Q)
             \, \delta^{(3)}(\vec p_{\overline H} - z_2 \vec p_{\overline Q})
   \label{dspADD}
\end{eqnarray}
where $z$ is the fraction of the parent quark momentum carried by the 
final-state meson. 

Quark fragmentation has been described by several different models.  A
fragmentation function smears the quark momentum into the momentum of
the final-state meson during hadronization.  Typical fragmentation
functions are a delta function or the Peterson function
\cite{PETERSON}. In this study we have used the delta function
but have also checked our results using the Peterson
function. Delta function fragmentation simply assumes that the quark
and meson momenta are identical,
\begin{equation}
  D^H_Q(z) = \delta(1-z) \, \, .
\end{equation}
On the other hand, the Peterson function assumes that 
the meson momentum is smeared according to
\begin{equation}
  D^H_Q(z) = \frac{N}{z(1-(1/z)-\epsilon_Q/(1-z))^2} \, \, .
\end{equation}
The normalization, $N$, is fixed by the requirement $\sum_H \int_0^1
dz\, D_Q^H(z)=1$.  The peak of the fragmentation function is at $z
\approx 1 - 2\epsilon_Q$ with a width $\epsilon_Q \approx (m_q/m_Q)^2$
\cite{PETERSON}.  We have assumed e.g.\ that all the $c$ quarks
fragment into $D$ mesons so that $H \equiv D$ for charm.

The hadronic heavy quark production cross section {\em per nucleon} in p$A$
collisions can be factorized into the general form
\begin{equation}
\frac{1}{A} E_{Q}E_{\overline Q} \frac{d\sigma^{{\rm p}A \rightarrow
   Q \overline Q + X}}
    {d^{3} \vec p_Q d^3 \vec p_{\overline Q}} = \sum_{i,j}
    \int_0^1 dx_1 \int_0^1 dx_2 \, f_i^{\rm p}(x_1,Q^2) f_j^A(x_2,Q^2) 
     E_{Q}E_{\overline Q}
    \displaystyle \frac{d\hat \sigma^{ij \rightarrow Q \overline Q}}
                       {d^3 \vec p_{Q} d^3 \vec p_{\overline Q}} \, \, .
      \label{dscc}
\end{equation}
The parton distributions $f_i^{\rm p}$ and $f_j^A$, described in
Eq.~(\ref{hardpA}), are evaluated at
the scale $Q^2 \sim m_{T_Q}^2$.  To 
lowest order, LO, the partonic cross section is
\begin{equation}
 \displaystyle E_{Q} E_{\overline Q}\frac{d\hat \sigma^{ij \rightarrow
 Q \overline Q}} {d^3 \vec p_{Q} d^3 \vec p_{\overline Q}}  =
  \frac{\shat}{2\pi}\frac{d\hat\sigma^{ij\rightarrow Q\overline Q}}{d\that}
  \delta^{(4)}(p_1+p_2-p_{Q}-p_{\overline Q}) \, \, ,
  \label{dsccLO}
\end{equation}
where $p_1$ and $p_2$ are the four momenta of the incoming partons. 
The lowest order convolution of the parton distributions with the 
subprocess cross section is written as
\begin{eqnarray}
  \displaystyle \lefteqn{\sum_{i,j} f_i^{\rm p}(x_1,Q^2) 
     f_j^A(x_2,Q^2) \frac{d\hat 
    \sigma^{ij \rightarrow Q \overline Q}}{d \that}
     = \displaystyle \frac{1}{16 \pi \shat^2}
         \bigg( f_g^{\rm p}(x_1,Q^2) f_g^A(x_2,Q^2) \overline{|{\cal M}_{gg 
      \rightarrow Q\overline Q} |}^2} &  & \nonumber \\
          &   & \mbox{} + \sum_{q=u,d,s} [f_q^{\rm p}(x_1,Q^2) f_{\overline 
      q}^A(x_2,Q^2) + f_{\overline q}^{\rm p}(x_1,Q^2) f_q^A(x_2,Q^2)] 
      \overline{|{\cal M}_{q\overline q \rightarrow Q\overline Q} |}^2 \bigg)
      \, \, .
     \label{sigmahat}
\end{eqnarray}
The squared matrix elements $\overline{|{\cal M}_{gg \rightarrow 
Q\overline Q} |}^2$ and $\overline{|{\cal M}_{q\overline q \rightarrow 
Q\overline Q} |}^2$ can be found in e.g.\ Ref.~\cite{COMBRIDGE}.

Integration of Eqs.~(\ref{dspAll}) and (\ref{dspADD}) over the total
phase space gives the normalization $\sigma^{{\rm p}A \rightarrow l
\overline l} = B^2 \sigma^{{\rm p}A \rightarrow H \overline H} = B^2
\sigma^{{\rm p}A \rightarrow Q \overline Q}$, where
$B=\Gamma_l/\Gamma$.  The total cross section for lepton pair
production is then equal to the total cross section of heavy quark pair
production multiplied by the square of the lepton branching ratio.
The following expressions for the double differential distribution and
the total cross section of $Q \overline Q$ production per nucleon will
be useful for our later discussion:
\begin{eqnarray}
\frac{d \tilde\sigma^{Q\overline Q}}{d^{3} \vec p_Q d^3 \vec p_{\overline Q}} 
& = & \frac{1}{E_{Q}} \frac{1}{E_{\overline Q}}
 \bigg[ \frac{1}{A} E_{Q}E_{\overline Q} \frac{d\sigma^{{\rm p}A \rightarrow
   Q \overline Q +X}}
    {d^{3} \vec p_Q d^3 \vec p_{\overline Q}} \bigg]
\, \, , \label{sigtildediff} \\
\tilde\sigma^{Q \overline Q}
& = & \int d^3\vec p_{Q} \, d^3\vec p_{\overline Q} \frac{d \tilde\sigma^{Q
      \overline Q}}{d^{3} \vec p_Q d^3 \vec p_{\overline Q}} \nonumber  \\
& = & \int dp_T^2 dy_1 dy_2 \, \sum_{ij} x_1 f_i^p(x_1,Q^2) x_2 f_j^A(x_2,Q^2)
      \frac{d\hat\sigma^{ij \rightarrow Q \overline Q}}{d\that} \, \, ,
\label{sigtildetot}
\end{eqnarray}
where $x_{1,2} = m_T/\sqrt{s}\,(e^{\pm y_1}+e^{\pm y_2})$ and $m_T^2 =
p_T^2+m_Q^2$. The rapidities of the heavy quarks are $y_{1,2}$. The
total $Q \overline Q$ cross sections per nucleon in pp and p$A$
interactions are presented in Table \ref{sigtable}.

%
%
\begin{table}
\begin{tabular}{cccc}
\hline
\hline
   \raisebox{0pt}[12pt][5pt]{}  & SPS ($A=208$) [$\mu$b] & RHIC ($A=197$) [$\mu$b] & LHC ($A=208$)  [$\mu$b]\\
\hline
  \raisebox{0pt}[12pt][5pt]{$\tilde\sigma_{{\rm pp}}^{c\overline c}$} 
& 1.188 & 143.7 & 6137 \\
   \raisebox{0pt}[13pt][5pt]{$\tilde\sigma_{{\rm p}A}^{c\overline c}$} 
& 1.282 & 140.1 & 4826 \\
   \raisebox{0pt}[12pt][7pt]{$\tilde\sigma_{{\rm pp}}^{b\overline b}$} 
& $1.674 \times 10^{-5}$ & 1.412 & 251.9 \\
  \raisebox{0pt}[13pt][7pt]{$\tilde\sigma_{{\rm p}A}^{b\overline b}$} 
& $1.464 \times 10^{-5}$ & 1.513 & 226.7 \\
\hline 
\hline
\end{tabular} 
\caption{The total $Q \overline Q$ production cross sections per nucleon
in pp and p$A$ interactions.}
\label{sigtable}
\end{table}

%
%

\subsection{Correlated and uncorrelated pairs in p$A$ collisions}

In p$A$ collisions, a number of uncorrelated lepton pairs are also
produced along with the correlated lepton pairs. Since, in most cases,
one would prefer to study either correlated or uncorrelated pairs
alone, these events must be separated. One method of separation is
like-sign subtraction because like-sign pairs are all uncorrelated: 
$l^+$'s originate from e.g.\ $c$ quarks whereas $l^-$'s come
from $\overline c$ quarks.  Another possible method is to use
azimuthal angle cuts. At lowest order, correlated $Q\overline Q$ pairs
are produced back-to-back.  Thus, leptons produced from these pairs
tend to have a larger relative azimuthal angle than uncorrelated
ones. With a suitable azimuthal cut the relative amount of
uncorrelated pairs could possibly be reduced. However, the focus of
our study is not how to separate the pairs but an estimate of their
rates.

In order to calculate the number of correlated and uncorrelated pairs,
we assume that all subprocess interactions are independent.  In this
case, the number of interactions can be described by a Poisson
distribution.  In our determination of the relative number of
correlated and uncorrelated pairs, it is convenient to consider the
impact-parameter integrated $Q \overline Q$ cross section,
\begin{equation}
  \sigma^{pA\rightarrow Q\overline Q + X} =
    \int d^2 {\bf b} \, (1-e^{-T_{A}({\bf b})\tilde\sigma^{Q\overline Q}})
   =   \int d^2 {\bf b} \, \sum_{N=1}^{\infty}
       \frac{\overline{N}_{Q\overline Q}^N({\bf b}) 
         e^{- \overline{N}_{Q\overline Q}({\bf b})}}{N!} \, \, ,
    \label{sigmapAcc}
\end{equation}
where $ {\bf b}$ is the impact parameter.  The number of $Q \overline Q$
pairs produced as a function of impact parameter is $\overline{N}_{Q \overline
Q}({\bf b}) = T_{A}({\bf b})\tilde\sigma^{Q\overline Q}$ with
$T_{A}({\bf b}) = \int dz \rho_A({\bf b},z)$ the thickness function of
a nucleus with density $\rho_A({\bf b},z)$.

Multiplying the right hand side of Eq.~(\ref{sigmapAcc}) by the unit normalized
$Q \overline Q$ cross section, 
\begin{equation} 1^N= \bigg( \frac{1}{\tilde\sigma^{ Q\overline Q}}\int d^3 
      \vec p_{Q} d^3 \vec p_{\overline Q} 
       \frac{d\tilde\sigma^{Q \overline Q}}
            {d^3 \vec p_{Q i} d^3 \vec p_{\overline Q i}} \bigg)^N \, \, ,
\label{unity}
\end{equation}
and differentiating with respect to 
$d^3 \vec p_{Q} d^3 \vec p_{\overline Q}$ we find
\begin{eqnarray}
 \displaystyle \frac{d \sigma^{{\rm p}A \rightarrow Q \overline Q + X}}
                    {d^3 \vec p_{Q} d^3 \vec p_{\overline Q}} & = & 
     \displaystyle
     \int d^2 {\bf b} \, \sum_{N=1}^{\infty}
        \frac{\overline{N}_{Q\overline Q}^N ({\bf b})
           e^{- \overline{N}_{Q\overline Q}({\bf b})}}{N!}
     \prod_{i=1}^{N} \Big( \frac{1}{\tilde\sigma^{Q\overline Q}} 
     \int d^3\vec p_{Q i} d^3\vec p_{\overline Q i} 
       \frac{d \tilde\sigma^{Q \overline Q}}{d^3\vec p_{Qi} d^3\vec 
     p_{\overline Q i}}\Big) 
     \nonumber \\
     &  & \displaystyle
        \times \Big(\sum_{j,k=1}^{N} 
          \delta^{(3)}(\vec p_{Q} -\vec p_{Q j})
          \delta^{(3)}(\vec p_{\overline Q} -\vec p_{\overline Q k}) \Big) \,
      \, .
      \label{sigmaABgeneral}
\end{eqnarray}
We have written the power $N$ in Eq.~(\ref{unity}) as a product in
order to separate the momenta $\vec p_{ i}$ of different
subcollisions. We have also written the delta functions as a sum over
$j$ and $k$ in order to indicate to which subcollision they
belong. When considering correlated pairs, both quarks come from the
{\em same} subcollision so that $j=k$.  In the case of uncorrelated
pairs, $j\neq k$, the $Q$ and $\overline Q$ are picked from different
subcollisions.

Thus for correlated pairs, $j=k$, we have
\begin{eqnarray}
  \displaystyle \frac{d \sigma^{{\rm p}A \rightarrow Q \overline Q +X}_{\rm corr}}
                    {d^3 \vec p_{Q} d^3 \vec p_{\overline Q}} &  = &
  \displaystyle \int d^2 {\bf b} \, \sum_{N=1}^{\infty}
        \frac{\overline{N}_{Q\overline Q}^N ({\bf b})
           e^{- \overline{N}_{Q\overline Q}({\bf b})}}{N!}
        N \, \frac{1}{\tilde\sigma^{Q \overline Q}} \,
        \frac{d\tilde \sigma^{Q \overline Q}}
             {d^3 \vec p_{Q} d^3 \vec p_{\overline Q}} \nonumber \\
  & = & \displaystyle 
      \int d^2 {\bf b} \,\, T_{A}({\bf b}) 
      \frac{d\tilde \sigma^{Q \overline Q}}{d^3 
      \vec p_{Q} d^3 \vec p_{\overline Q}} \, \, .
      \label{correlated}
\end{eqnarray}
Similarly, for uncorrelated pairs, $j\neq k$,
\begin{eqnarray}
  \displaystyle \frac{d \sigma^{{\rm p}A \rightarrow Q \overline Q+X}_{\rm 
   uncorr}}
                    {d^3 \vec p_{Q} d^3 \vec p_{\overline Q}} & = &
  \displaystyle \int d^2 {\bf b} \, \sum_{N=1}^{\infty}
        \frac{\overline{N}_{Q\overline Q}^N ({\bf b})
           e^{- \overline{N}_{Q\overline Q}({\bf b})}}{N!}
        N(N-1) \, 
        \frac{1}{\tilde\sigma^{Q \overline Q}} \,
        \frac{d\tilde \sigma^{ Q \overline Q}}
             {d^3 \vec p_{Q}}       \,
        \frac{1}{\tilde\sigma^{Q \overline Q}} \,
        \frac{d \tilde\sigma^{ Q \overline Q}}
             {d^3 \vec p_{\overline Q}} \nonumber \\
  & = & \displaystyle 
      \int d^2 {\bf b} \,\, T_{A}^{\,\,2}({\bf b}) 
      \frac{d\tilde \sigma^{Q \overline Q}}{d^3 \vec p_{Q} }
      \frac{d \tilde \sigma^{Q \overline Q}}{d^3 \vec p_{\overline Q}} \, \, .
      \label{uncorrelated}
\end{eqnarray}
We use the definition of the nuclear overlap function,
$T_{AB}({\bf b}) = \int d {\bf s} \, 
  T_{A}({\bf s}) T_{B}({\bf b}-{\bf s})$,
with $A=B$ to obtain the integral over ${\bf b}$ in Eq.~(\ref{uncorrelated}),
$\int d^2{\bf b} \,  T_{A}^{\,2}({\bf b}) =  
  T_{AA}({\bf 0}) = A^2/(\pi R_A^2)$.
The last equality is obtained by assuming diffuse Woods-Saxon
distributions for the nuclear density. Thus, for correlated
pairs we find
\begin{equation}
  \displaystyle \frac{d \sigma^{{\rm p}A \rightarrow Q \overline 
     Q+X}_{\rm corr}}
                    {d^3 \vec p_{Q} d^3 \vec p_{\overline Q}}  = 
      A \frac{d\tilde \sigma^{Q \overline Q}}{d^3 \vec p_{Q} 
       d^3 \vec p_{\overline Q}} \, \, ,
      \label{corrcalc}
\end{equation}
while for uncorrelated pairs
\begin{equation}
  \displaystyle \frac{d \sigma^{{\rm p}A \rightarrow Q \overline Q+X}_{\rm
     uncorr}}
                    {d^3 \vec p_{Q} d^3 \vec p_{\overline Q}}  = 
      \frac{A^2}{\pi R_A^2}
      \frac{d\tilde \sigma^{Q \overline Q}}{d^3 \vec p_{Q} }
      \frac{d\tilde \sigma^{Q \overline Q}}{d^3 \vec p_{\overline Q}} \, \, .
      \label{uncorrcalc}
\end{equation}
The double differential cross section for the pairs, the right-hand
side of Eq.~(\ref{corrcalc}), was defined in
Eq.~(\ref{sigtildediff}). The single $Q$ cross section, $d \tilde
\sigma^{Q \overline Q}/d^3 \vec p_{Q}$, is the integral of
Eq.~(\ref{sigtildediff}) over $d^3 \vec p_{\overline Q}$.  Dividing
Eqs. (\ref{corrcalc}) and (\ref{uncorrcalc}) by $A$ gives the
$Q\overline Q$ production cross section per nucleon.

The above formalism can be extented to $AB$ collisions by simply
replacing the thickness function $T_A$ in Eq. (\ref{sigmapAcc}) by the
nuclear overlap function $T_{AB}$.

As seen in Eq.~(\ref{uncorrcalc}), the uncorrelated pair cross section
is proportional to the product of two independent single heavy quark
cross sections.  The heavy quark cross sections are, in turn,
proportional to the lepton cross sections.  This proportionality to
the lepton cross sections could be utilized to study the ratio of
cross sections of uncorrelated lepton pairs from p$A$ and pp
collisions. This uncorrelated ratio would probe the square of the
nuclear modifications of gluon distribution function, $({R_{g}^A})^2$,
instead of $R_{g}^A$.  Because the uncorrelated pair cross section is
the product of independent single lepton distributions, lepton pairs
produced by leptons from two different nucleon-nucleon collisions can
be used to simulate uncorrelated pairs.

%
%

\section{Numerical inputs}

We calculate the lepton pair cross section in Eq.~(\ref{dspAll}) by
generating lepton pairs in proton-nucleus collisions by Monte Carlo
and binning them according to their mass and rapidity.  We calculate
$\langle x_1 \rangle$, $\langle x_2 \rangle$ and $\langle Q \rangle$
in each $(y_{l \overline l},M_{l \overline l})$ bin.  The $\langle x_2
\rangle$ regions probed are indicated in Fig.~\ref{RGDIST}.  For
simplicity, since isospin effects do not play a role in $gg
\rightarrow Q\overline Q$, we have assumed that the nuclei consist of
$A$ protons.  We have used the MRST LO (central-gluon) parton
distribution functions \cite{MRST98} from PDFLIB \cite{PDFLIB} along
with the EKS98 \cite{EKS98} nuclear modifications. The quark masses
are $m_c=1.2$ GeV and $m_b=4.75$ GeV while the meson masses are $m_D=
1.8693$ GeV and $m_B= 5.278$ GeV.  In $c\overline c$ production we
have chosen the factorization scale to be twice the quark transverse
mass, $Q^2=4m_{T_c}^2$, because the lowest values of $m_{T_c}^2$ might
otherwise be below the minimum $Q^2$ of the parton distribution
set. Since the transverse mass is larger in $b\overline b$ production,
we use $Q^2=m_{T_b}^2$.

The $c$ quark fragments into $D^+$, $D^0$ and $D_s$ mesons with
respective relative weights 1:1:0.3 \cite{PYTHIA}. Here we neglect
$D_s$ mesons and allow $c$ quarks to only fragment to $D^+$ or to
$D^0$ with equal probabilities which should not significantly affect
the results.  Similarly: $\overline c \rightarrow D^-$ and $\overline
D^0$; $b \rightarrow B^+$ and $\overline B^0$; and $\overline b
\rightarrow B^-$ and $B^0$.

In order to obtain sufficiently high statistics, we generated $2
\times 10^6$ $Q \overline Q$ pairs and let each pair fragment $10^3$
times.  We used JETSET/PYTHIA subroutines \cite{PYTHIA} for meson
decays to leptons. To further improve statistics and reduce the
runtime of the calculation, we allowed leptonic decays only. The final
normalization is obtained using the leptonic branching ratios. The
total $Q\overline Q$ production cross sections per nucleon needed for
normalization are given in Table 1.
 
We also generated uncorrelated lepton pairs from two $Q \overline Q$
pairs which then separately fragment and decay.  The lepton pairs
originating from different quark pairs are the uncorrelated ones.
This is the origin of the factor of two in the previous paragraph.

We then binned all correlated and uncorrelated $e^\pm e^\mp$, $e^\pm
\mu^\mp$, and $\mu^\pm \mu^\mp$ pairs.  We accept only leptons which
originate directly from $D \overline D$ and $B \overline B$ meson
decays and neglect the background from secondary or tertiary leptons
from e.g.\ $D^+ \rightarrow K^+ X \rightarrow \mu^+ X'$.

Our pp and p$A$ results are calculated for SPS ($\sqrt{s}=$17.3 GeV),
RHIC ($\sqrt{s}=$200 GeV) and LHC ($\sqrt{s}$=5500 GeV), using the
same center of mass energies as in $AA$ collisions to probe the same
$x$ values in all collisions.  At the SPS and LHC, we assume a Pb
($A$=208) target and an Au ($A$=197) target for RHIC.  The new NA60
\cite{NA60} experiment at the SPS makes it possible to detect muon
pairs in the range $0\leq \eta_\mu \leq 1$.  Since the low energy
makes $b \overline b$ production negligible, only $c\overline c$
production is studied at the SPS.  At RHIC, the PHENIX detector
\cite{PHENIX} is designed to measure $ee$, $e\mu$ and $\mu\mu$
pairs. The detector has two electron and two muon arms.  The
acceptance of electron arms is $-0.35 < \eta_e < 0.35$ in
pseudorapidity and $\pm(22.5^{\circ} \leq \phi_e \leq 112.5^{\circ})$
in azimuth. The forward muon arm covers the pseudorapidity region
$1.15 < \eta_{\mu} <2.44$ with almost full azimuthal acceptance. The
backward muon arm has a similar acceptance.  We only consider leptons
from the forward arm so as not to mix regions of $x_2$.  The ALICE
\cite{ALICE} detector is designed to study nucleus-nucleus
interactions at the LHC. They will measure electrons at central
rapidity, $|\eta_e| \leq 0.9$, and muons in a forward muon
spectrometer with $2.5 \leq \eta_\mu \leq 4$.  We impose the cuts
described above on the lepton pseudorapidity in order to simulate
detector acceptances.  These are summarized in Table \ref{ycuts}.
However, no cuts on minimum lepton energy or azimuthal angle were
imposed.

%
\begin{table}
\begin{center}
\begin{tabular}{cccc}
\hline
\hline
          & SPS      & RHIC         & LHC         \\

\hline
 electron & $0.0 \leq \eta_e \leq 1.0$ &  $-0.35 \leq \eta_e \leq 0.35$ & 
$-0.9 \leq \eta_e \leq 0.9$   \\
 muon     & $0.0 \leq \eta_\mu \leq 1.0$ &  $1.15 \leq \eta_\mu \leq 2.44$
 & $2.5  \leq \eta_\mu \leq 4.0$  \\
\hline
\hline 
\end{tabular}
\end{center}
\caption{The cuts on the pseudorapidity of single leptons for the three
energy regimes studied are indicated both for electrons and muons. }
\label{ycuts}
\end{table}

%
%

\section{Results}

\begin{figure}[!t]
\centering
\includegraphics[width=11cm]{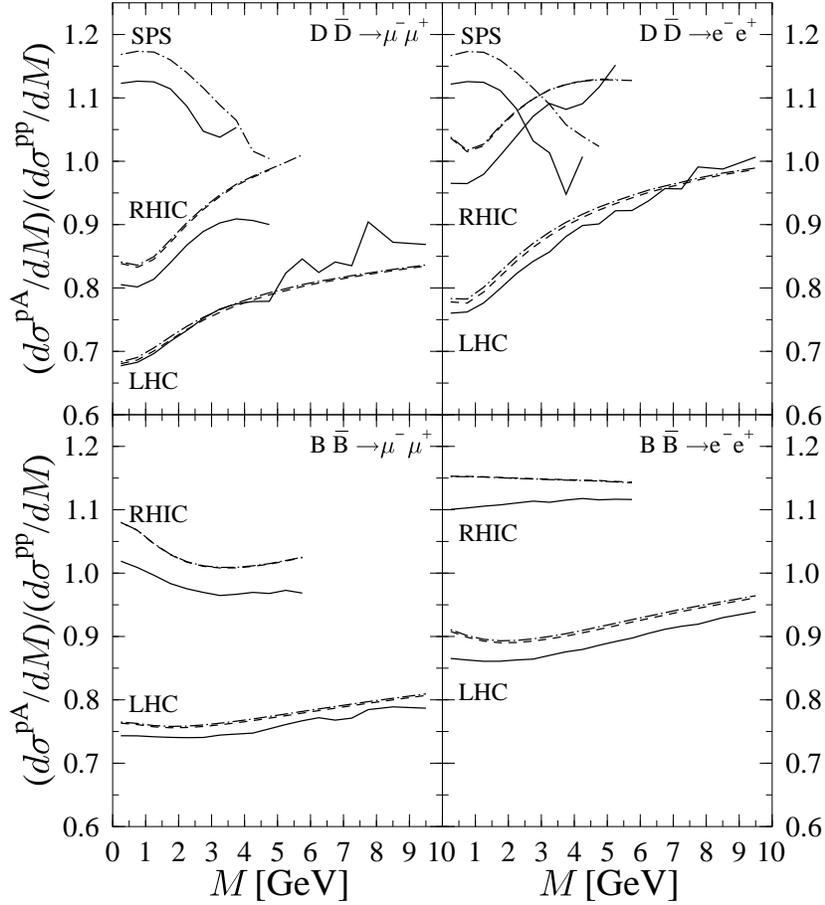}
\caption[a] { {\small The ratio of $e^\pm e^\mp$ and $\mu^\pm \mu^\mp$
pair cross sections from correlated $D \overline D$ and $B \overline
B$ decays in p$A$ and pp collisions, $r_\sigma$, defined in
Eq.~(\ref{rsigdef}), at SPS, RHIC and LHC energies (solid curves). The
nuclear gluon distribution, $R_g^A(x_2,Q)$, at $\langle x_2 \rangle$
and $\langle Q \rangle$ ($\sqrt{\langle Q^2 \rangle}$) of each 
$M_{l \overline l}$ bin is indicated by the dashed (dotted-dashed) 
curves.
}}
\label{FIGDSRG}
\end{figure}

We begin by comparing the ratios of lepton pair cross sections with
the input $R_g^A$.  Figure \ref{FIGDSRG} shows the ratio of
differential cross sections of $e^\pm e^\mp$ and $\mu^\pm \mu^\mp$
pairs from correlated $D \overline D$ and $B \overline B$ decays in
p$A$ and pp collisions at the SPS, RHIC and LHC energies as a function
of the lepton pair invariant mass (solid curves). These results are
compared to the ratios $R_g^A(\langle x_2 \rangle, \langle Q \rangle)$
(dashed curves) and $R_g^A(\langle x_2 \rangle, \sqrt{\langle Q^2
\rangle})$ (dotted-dashed curves) at the average $x_2$ and $Q$ of each
mass bin summed over $y_{l \overline l}$.  All the results are
integrated over the rapidity interval given in Table~\ref{ycuts}.  For
brevity, we define the ratio of the cross sections as a function of
lepton pair mass as
\begin{equation}
r_{\sigma} \equiv \frac{d\sigma^{{\rm p}A}/dM}{d\sigma^{{\rm pp}}/dM} \, \, .
\label{rsigdef}
\end{equation}
Note that $r_\sigma$ follows $R_g^A$ to a very good approximation at all
energies. The higher the energy, the better the agreement: at LHC
$r_{\sigma}$ agrees with $R_g^A$ very well. Note also that $R_g^A(\langle x_2
\rangle, \langle Q \rangle) \approx R_g^A(\langle x_2 \rangle,
\sqrt{\langle Q^2 \rangle})$ in all cases.

The results shown in Fig.~\ref{FIGDSRG} are obtained using a delta
function for the fragmentation. If the Peterson function is used with
$\epsilon_c=0.15$ and $\epsilon_b=(m_c/m_b)^2\epsilon_c$
\cite{PETERSON} , the outcome is very similar.  The difference in
$r_\sigma$ due to the fragmentation function is smaller than or on the
order of the statistical fluctuations in the Monte Carlo.

%
%

The ratio $r_{\sigma}$ always lies below $R_g^A$ for two reasons.
First, some fraction of $Q \overline Q$ production is through the $q
\overline q \rightarrow Q\overline Q$ channel.  Quark shadowing
effects are also included in the calculation of $r_\sigma$.  Second,
the phase space integration smears the shadowing effect relative to
$R_g^A(\langle x_2 \rangle, \langle Q \rangle)$.  The $q
\overline q$ contribution decreases with energy while $R_g^A$ changes
more slowly over the $x_2$ region at the LHC, leading to better
agreement at this energy.

We have studied the influence of the $q \overline q$ channel by
turning off the quark distributions in our simulation, i.e.\ we fix
$f_q^A = f_{\overline q}^A = 0$ for all quark flavours.  The resulting
ratio $r_{\sigma}(f_q^A=0)$ always lies above $r_{\sigma}$, indicating
that quark effects reduce the cross section ratios. However,
$r_{\sigma}(f_q^A=0)$ is still below $R_g^A$, suggesting that other
effects influence $r_\sigma$. For example, at the SPS with $M=1.5$
GeV, $r_{\sigma} \approx 1.12$ and $r_{\sigma}(f_q^A =0) \approx 1.16$
whereas $R_g^A \approx 1.19$.

In order to estimate the effect of the phase space integration, we
reformulate the ratio of lepton pair cross sections as
\begin{equation}
  r_{\sigma} 
  = \frac{ \displaystyle\int d({\rm PS})_{Q\overline Q} 
           \,R_g^A(x_2,Q) h(x_1,x_2,m_T,\ldots)}
         { \displaystyle\int d({\rm PS})_{Q\overline Q} 
           \,h(x_1,x_2,m_T,\ldots)},
           \label{crratio}
\end{equation}
where $h(x_1,x_2,m_T,\ldots)$ is a complicated function which contains the
$Q\overline Q$ production cross section, fragmentation to mesons, and 
the subsequent meson decay to
leptons.  The integration over the $Q \overline Q$ phase space is denoted
by $d({\rm PS})_{Q\overline Q}$.  We define the average values
\begin{equation}
  \langle x_2 \rangle = \int d({\rm PS})_{Q\overline Q} \, x_2
  \,h(x_1,x_2,m_T,\ldots),\quad \langle Q \rangle = \int d({\rm
  PS})_{Q\overline Q} \, Q \,h(x_1,x_2,m_T,\ldots),
\end{equation}
and expand $R_g^A$ as a Taylor series around $\langle x_2 \rangle$ and
$\langle Q \rangle$ up to second order.  The terms linear in $(x_2 -
\langle x_2 \rangle)$ and $(Q - \langle Q \rangle)$ are zero after
integration and the ratio $r_{\sigma}$ becomes
\begin{eqnarray}
r_{\sigma} & = & R_g^A(\langle x_2 \rangle,\langle Q \rangle) 
   + \frac{1}{2}\frac{\partial^2 R_g^A(\langle x_2 \rangle,\langle Q 
     \rangle)}{\partial x_2^2} 
    \langle ( x_2 - \langle x_2 \rangle )^2 \rangle  \nonumber \\
 &  & \mbox{} + 
     \frac{1}{2}\frac{\partial^2 R_g^A(\langle x_2 \rangle,\langle Q 
     \rangle)}{\partial Q^2} 
     \langle(Q - \langle Q \rangle)^2 \rangle \nonumber \\
 &  &  + \frac{\partial^2 R_g^A(\langle x_2 \rangle,\langle Q 
     \rangle)}{\partial x_2 \, \partial Q} 
     \langle ( x_2 - \langle x_2 \rangle )
     (Q - \langle Q \rangle) \rangle + \dots
         \, \, . \label{expansion}
\end{eqnarray} 
The sum of the correction terms is always negative, thus the
ratio $r_{\sigma}$ always lies below $R_g^A$. A study of the partial
derivatives (see Fig.~1) shows that ${\partial^2 R_g^A}/{\partial
x_2^2} < 0$ in the vicinity of $\langle x_2\rangle $ for each energy
so that the first correction term is always negative. Near $\langle Q
\rangle$, we have ${\partial^2 R_g^A}/{\partial Q^2} > 0$ at SPS but $< 0$
at RHIC and LHC. The first correction term dominates
at SPS but the second one becomes dominant at the LHC energy. The net
deviation of $r_{\sigma}$ from $R_g^A(\langle x_2 \rangle,\langle Q
\rangle)$ is then a sum of different terms reflecting both $x$-
and $Q$-dependence of $R_g^A$. As also seen in Fig.~\ref{FIGDSRG}, the
ratio $r_{\sigma}$ deviates slightly more from $R_g^A(\langle
x_2\rangle,\langle Q\rangle)$ for the electron pairs than for the muon
pairs. The curvature of $R_g^A$ with
$x$ is stronger at larger values of $x$ and due to the differences in rapidity
coverage, the average values of $x_2$ are larger for
electron pairs.

%
%
\begin{figure}[!t]
\centering
\includegraphics[width=10cm]{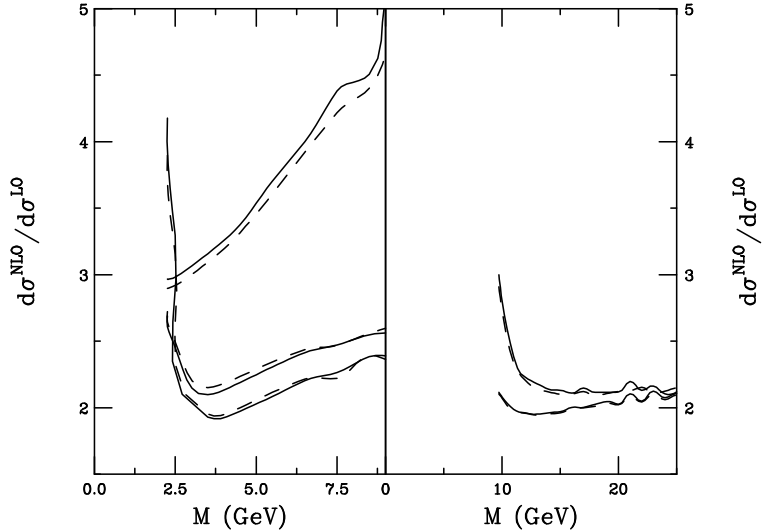}
\caption[a] { {\small The ratio of the NLO to LO differential cross
sections for $Q \overline Q$ production as a function of the invariant
mass of the $Q \overline Q$ pair.  Left panel: $c \overline c$ pairs
produced in p$A$ (solid) and pp (dashed) collisions at SPS (upper
curves), RHIC (middle curves) and LHC (lower curves) energy.  Right
panel: The same for $b \overline b$ production at RHIC (upper curves)
and LHC (lower curves). }}
\label{FIGKFAC}
\end{figure}

In order to estimate next to leading order (NLO) effects, we have also
calculated the differential cross section of $Q \overline Q$
production to NLO \cite{HPC}.  The results are shown in
Fig.~\ref{FIGKFAC} where the ratio $K = (d\sigma^{\rm
NLO}/dM)/(d\sigma^{\rm LO}/dM)$ is given for p$A$ and pp
collisions. This $K$ factor increases with $M$ but decreases with
energy. The figure also shows that the $K$ factor for p$A$ collisions,
$K_A$, is approximately equal to that for pp collisions, $K_{\rm p}$,
since
\begin{equation}
  \frac{d\sigma^{\rm NLO}_{{\rm p}A}}{d\sigma^{\rm LO}_{{\rm p}A}}
  \approx \frac{d\sigma^{\rm NLO}_{{\rm pp}}}{d\sigma^{\rm LO}_{{\rm pp}}}
  \Rightarrow 
  \frac{K_Ad\sigma^{\rm LO}_{{\rm p}A}}{d\sigma^{\rm LO}_{{\rm p}A}}
  \approx \frac{K_{\rm p} d\sigma^{\rm LO}_{{\rm pp}}}{d\sigma^{\rm 
   LO}_{{\rm pp}}}
  \Rightarrow
   K_A \approx K_{\rm p} \, \, . \label{kaeqkp}
\end{equation}
This result can be expected even though the $K$ factor is large
because the $gg$ channel dominates both LO and NLO $Q \overline Q$
production.  The $x_2$ regions probed at LO and NLO are also very
similar.  Equation (\ref{kaeqkp}) further implies that the ratio of
the p$A$ to pp cross sections at LO and NLO are approximately equal:
\begin{equation}
 \frac{d\sigma^{\rm NLO}_{{\rm p}A}}{d\sigma^{\rm NLO}_{{\rm pp}}}
 \approx\frac{d\sigma^{\rm LO}_{{\rm p}A}}{d\sigma^{\rm LO}_{{\rm pp}}}\, \, . 
\end{equation}
Thus, the NLO corrections to the ratios in Fig.~\ref{FIGDSRG} should
be small.  A similar result was found for Drell-Yan production at low
masses \cite{ekkv,EKST}. 

As explained earlier, the uncorrelated pairs might be used to obtain
the ratio $({R_g^A})^2$.  If the nuclear effects turn out to be too
small to be measured, that is, if $R_g^A \sim 1$, the squared ratio
might still be observable. We test whether the ratio of uncorrelated
lepton pair cross sections, $r_{\sigma}^{\rm uncorr}$ agrees with
$({R_g^A})^2$ at the LHC where the probability of uncorrelated
production is largest. Based on our previous discussion of the $K$ factor,
the NLO effects in the ratio $r_{\sigma}^{\rm uncorr}$ are expected
to remain small.

\begin{figure}[!t]
\centering
\includegraphics[width=9cm]{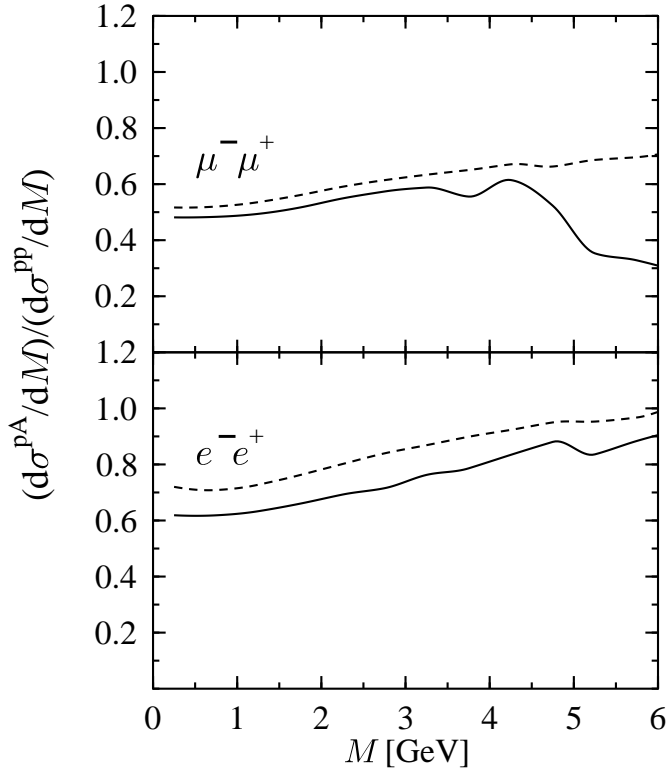}
\caption[a] { {\small Comparison of $[R_g^A(\langle x\rangle,\langle
Q\rangle)]^2$ (dashed) and uncorrelated lepton pair cross sections,
$r_\sigma^{\rm uncorr}$, for $D \overline D$ pair decays (solid) at
the LHC.}}
\label{Rg2uncorr}
\end{figure}

The results for $D \overline D$ decays at the LHC are shown in
Fig.~\ref{Rg2uncorr}.  It appears that the ratio of cross sections
does follow the slope of $[{R_g^A}(\langle x \rangle, \langle Q
\rangle)]^2$. However, the difference between the ratios is slightly
larger than in the correlated case. The origin of this difference can
be understood by expanding two $R_g^A$ factors in the ratio of uncorrelated
cross sections, $r_{\sigma}^{\rm uncorr}$, as in Eq.~(\ref{expansion}), and
noting that the deviations $\langle Q^2 \rangle - \langle Q
\rangle^2$ and $\langle x_2^2 \rangle - \langle x_2 \rangle^2$ are
larger for uncorrelated pairs.  Integration over the rapidity of one
of the leptons in a correlated pair increases the allowed phase space
and thus the rapidity range of the remaining lepton. This greater
available phase space directly increases the variation of $x_2$ and
$Q$ for uncorrelated pairs, leading to the larger deviations between
$r_\sigma^{\rm uncorr}$ and $({R_g^A})^2$.

\begin{figure}[!t]
\centering
\includegraphics[width=7cm]{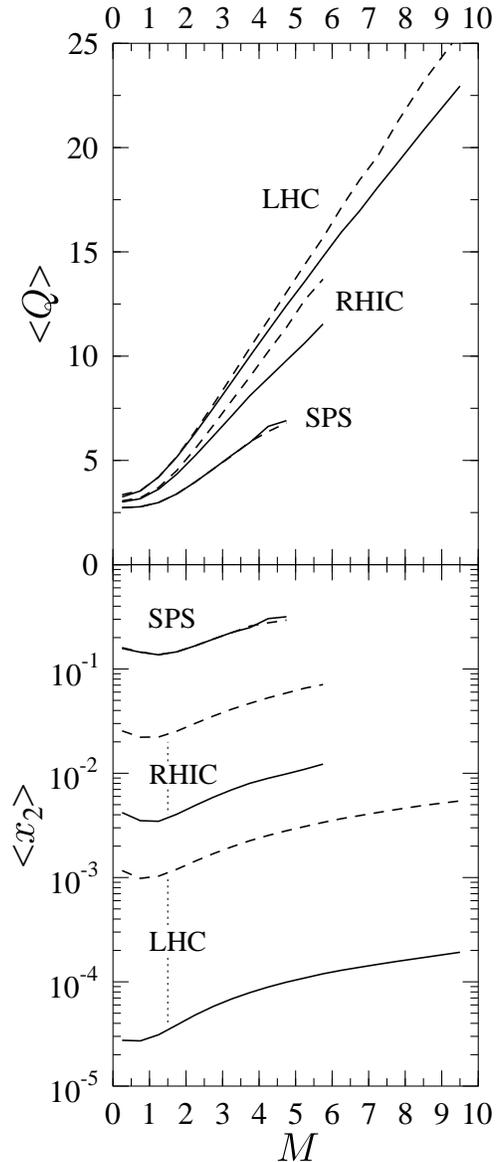}
\caption[a] { {\small We give $\langle x_2 \rangle$ and $\langle Q
\rangle$ as a function of pair mass $M$ for $D \overline D$ decays
at the three energies. The solid lines are for $\mu^-\mu^+$ pairs while the
dashed is for $e^-e^+$ pairs.  }}
\label{xQ2vsM}
\end{figure}

The slopes of the ratios $r_{\sigma}$ and $R_g^A$ (see
Fig.~\ref{FIGDSRG}) are determined by the momentum fractions probed by
the interaction, as seen in a comparison with Fig.~\ref{RGDIST}. 
The average momentum fraction of the parton
from nucleus, $\langle x_2 \rangle$, with muon pairs is 
$0.14 \leq \langle x_2 \rangle
\leq 0.32$ at the SPS, indicated in Fig.~\ref{RGDIST}. In this $x$
region, the ratio $R_g^A(x,Q^2)$ is decreasing.  At RHIC, the range of
momentum fractions probed is $0.003 \leq \langle x_2 \rangle \leq
0.012$, where the ratio is increasing quite rapidly. Finally, at the
LHC, $3\times 10^{-5} \leq \langle x_2 \rangle \leq 2\times 10^{-4}$
where $R_g^A(x,Q^2)$ is almost independent of $x$. The values of
$\langle x_2 \rangle$ are typically larger for electron pairs at
collider energies because the electron coverage is more central than
the muon coverage, as seen in Fig.~\ref{xQ2vsM} and Table~\ref{ycuts}.

%
%

We note here that the behaviour of $R_g^A(x,Q^2)$ depends not only on
$x_2$ but also on $Q^2$. Therefore the slope of $R_g^A(\langle
x\rangle,\langle Q^2\rangle)$, seen in Fig.~\ref{FIGDSRG}, does not
directly follow any of the curves shown in Fig.~\ref{RGDIST} but
evolves with $Q^2$ from one curve to another.  The average values 
$\langle Q^2 \rangle$ of $c\overline c$ decays are $7.58 \leq
\langle Q^2 \rangle \leq 48.5$ GeV$^2$ at the SPS, $9.46 \leq \langle Q^2
\rangle \leq 141$ GeV$^2$ at RHIC, and $11.4 \leq \langle Q^2 \rangle
\leq 577$ GeV$^2$ at the LHC.  The corresponding values of $\langle Q
\rangle$ are shown in Fig.~\ref{xQ2vsM}.  The minimum average
$Q^2$ are larger for $b \overline b$ production, $32.0 \leq \langle Q^2
\rangle \leq 54.3$ GeV$^2$ at RHIC and $37.9 \leq \langle Q^2 \rangle \leq
156$ GeV$^2$ at LHC.

%
%

The lepton pair cross sections from $B\overline B$ decays are also
shown in Fig.~\ref{FIGDSRG}.  The results are quite similar to those
from $D\overline D$ decays: at RHIC, the difference between
$r_{\sigma}$ and $R_g^A$ is $\sim 5$\% but decreases to 
$1-2$\% at the LHC.  The
slopes of these curves are different than for $D\overline D$,
they are almost constant with $M$. This constancy is due to the much
narrower $x$ and higher $Q^2$ regions probed by $b \overline b$
production. In a narrow $x$ region, the change in slope with $x$ is
small while at high $Q^2$, evolution is slower, as seen in
Fig.~\ref{RGDIST}.

%
%
\begin{figure}[!t]
\centering
\includegraphics[width=6.5cm]{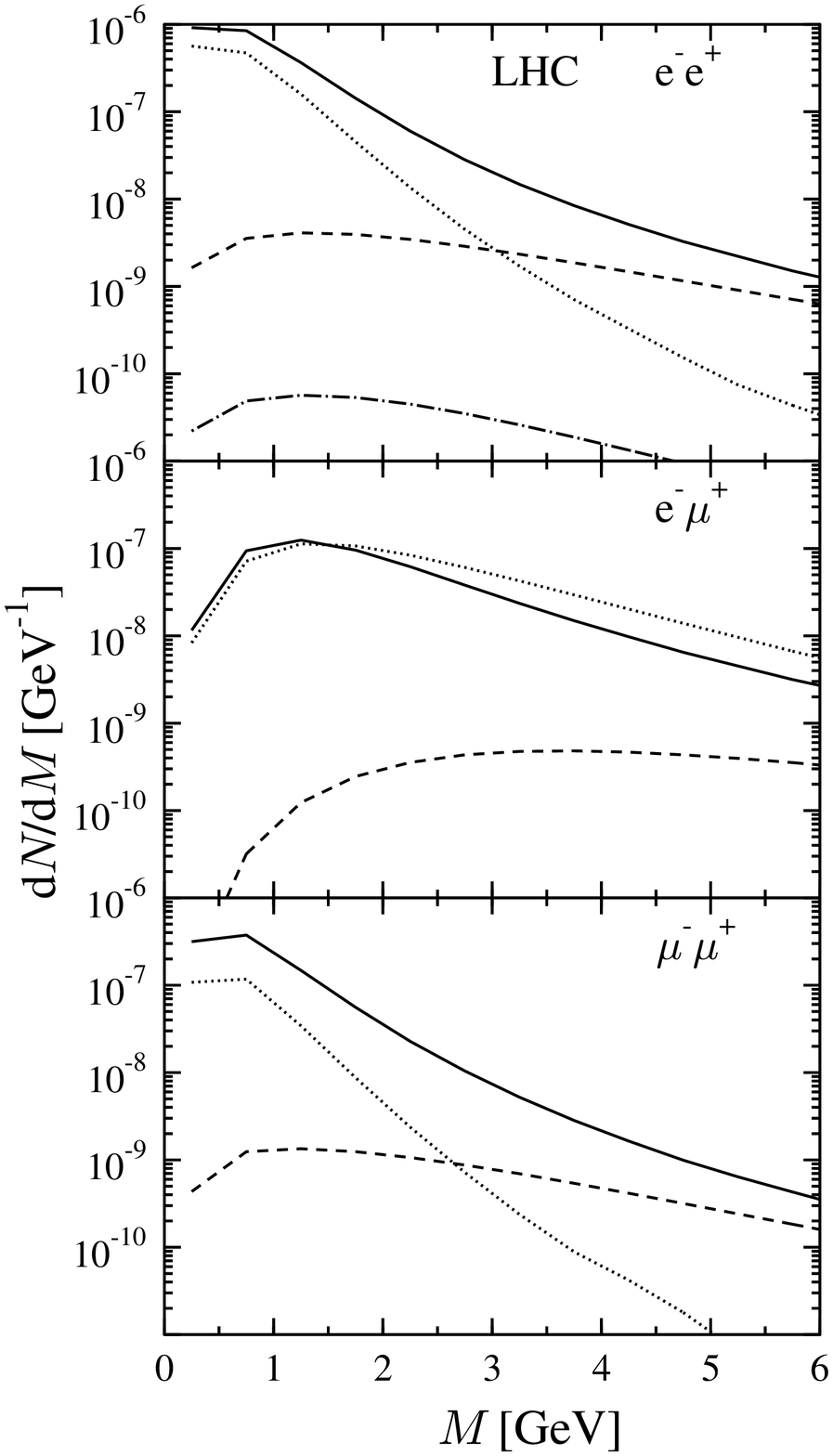}
\includegraphics[width=6.5cm]{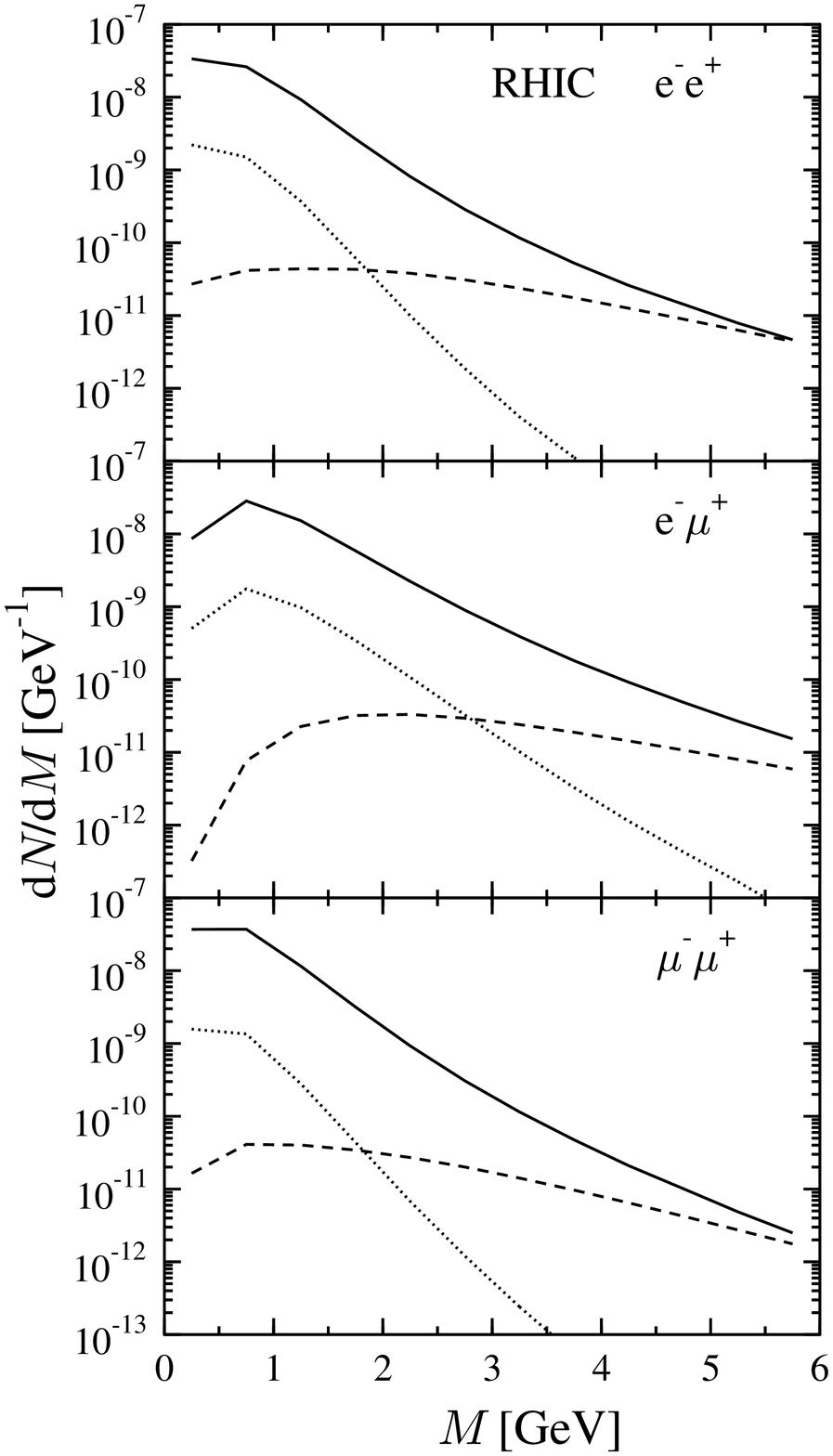}
\caption[a] { {\small Comparison of lepton pairs from
correlated and uncorrelated $D \overline D$ and $B \overline B$ pairs at LHC
(left) and RHIC (right).  The differential number
densities of $e^\pm e^\mp$, $e^\pm \mu^\mp$ and $\mu^\pm \mu^\mp$ pairs 
are given as a function
of $M$. The lepton pairs originate from correlated $D \overline D$
(solid), correlated $B \overline B$ (dashed), uncorrelated $D \overline D$
(dotted) and uncorrelated $B \overline B$ (dot-dashed) decays. The results are
shown without $K$ factors. }}
\label{FIGDNM}
\end{figure}

Figure \ref{FIGDNM} shows the mass dependence of the differential
number density distribution of lepton pairs in central collisions,
obtained from Eqs. (\ref{correlated}) and (\ref{uncorrelated}) through
differentiation with respect to ${\bf b}$ and setting ${\bf b} = 0$.
These results are shown without any $K$ factors, so that the final
result will be shifted when a $K$ factor is applied. In particular,
the relative normalization of the curves will change since the
correlated pairs will be multiplied by $K$ while the uncorrelated
pairs are increased by $K^2$.  Taking this correction into account,
one sees that at the LHC the number of uncorrelated pairs at $M \gsim
2-3$ GeV is of the same order as the correlated ones. At higher $M$,
the uncorrelated pairs are reduced and can be safely neglected, except
in the $e \mu$ configuration where the uncorrelated pairs are
larger. On the other hand, lepton pairs originating from $B \overline
B$ decays are negligible at low $M$ but become important at $M \gsim
5-6$ GeV. At RHIC, leptons from correlated $D \overline D$ decays
clearly dominate those from uncorrelated pairs over all masses so that
the uncorrelated pairs can be neglected. Note that at this energy
also, leptons from $B \overline B$ decays are of the same magnitude as
the correlated $D \overline D$ decays for $M \gsim 4-5$ GeV.

In Ref.~\cite{GGRV96-CHARM} it was shown that, in $AA$ collisions,
lepton pairs from uncorrelated heavy quark decays tend to have higher
masses than those from correlated decays.  This is because the
rapidity gap, $\Delta y$, between the quarks is larger when they are
uncorrelated and $M^2 \propto 1 + \cosh \Delta y$.  The uncorrelated
rate is much smaller in p$A$ interactions, however, which would reduce
the probability of a large $\Delta y$ relative to $AA$ collisions.
Including a finite rapidity cut severely reduces the number of leptons
from uncorrelated decays falling within the acceptance.  The effect of
the rapidity cut is to reduce the average $M$ for uncorrelated
relative to correlated pairs.  Thus in Fig.~\ref{FIGDNM} lepton pairs
from uncorrelated decays are most strongly reduced at large masses
when the rapidity acceptance is small, as in the $e^+ e^-$ coverage at
RHIC.  Since the $e \mu$ measurement involves mid-rapidity electrons
and forward (or backward) muons, the rapidity coverage is largest and
fewer uncorrelated decays are rejected.  In fact, the larger $e \mu$
coverage at the LHC as well as the higher $D \overline D$ rates lead
to more uncorrelated pairs at larger masses than correlated pairs.

The results in Fig.~\ref{FIGDNM} are for opposite-sign lepton pairs
only.  The uncorrelated yield can be determined using a like-sign
subtraction since like-sign pairs are all trivially uncorrelated. The
number of uncorrelated opposite-sign pairs is equivalent to the number of
like-sign pairs so it is relatively easy to estimate the real amount
of uncorrelated pairs. After this estimate, the uncorrelated pairs can
be used to study $({R_g^A})^2$, as shown in Fig.~\ref{Rg2uncorr}.

Finally we wish to mention that along with ``normal'' uncorrelated
lepton pairs originating from $D \overline D$ and $B \overline B$
decays a number of ``crossed events'', i.e.\ uncorrelated leptons from
$D \overline B$ and $B \overline D$ decays, are also produced. Their
rate lies between those from uncorrelated $D \overline D$ and $B
\overline B$ pairs. The total number of observed lepton pairs from
heavy quark decays is the sum of all pairs from $D\overline D$,
$B\overline B$, $D\overline B$ and $\overline D B$ decays. It is
necessary to know the relative number of pairs produced in each case
in order to decompose the results, as in Fig.~\ref{FIGDNM}.  However,
since uncorrelated $B \overline B$ pairs are generally negligible, we
can also neglect these ``crossed events'' and assume that it is
possible to deduce which lepton pairs came from correlated $D
\overline D$ pairs.  We do not consider any ``crossed events'' from
$D$ or $B$ decays combined with leptons from $\pi$ or $K$ decays.
These should predominantly appear at lower masses and can be removed
by like-sign subtraction.

%
%

\section{Conclusions}

We have calculated lepton pair production from heavy quark decays in
p$A$ and pp collisions using the EKS98 nuclear modifications to the
parton distributions. We have shown that the resulting cross section
ratios of p$A$ to pp rates reflect the initial nuclear modifications
of gluons rather well, as seen in Fig.~\ref{FIGDSRG}. The deviations
caused by phase space and the finite quark contributions are
analyzed. Based on these results, we conclude that if the ratio of
leptonic cross sections were measured within an acceptable accuracy,
the nuclear gluon distributions could be determined.  Several such
measurements have been proposed.

The NA60 experiment \cite{NA60} at the SPS plans to measure charm
production through lepton decays.  At this energy, a 20\%
antishadowing in $R_g^A$ manifests itself as a 10\% enhancement in the
ratio of cross sections due to the finite quark component. Assuming
the measurements could be done to $\sim 10$\% accuracy, NA60 results
could be used to pinpoint the behaviour of nuclear gluon distribution
in the range $0.17 \lsim x \lsim 0.4$.

At higher energies, the $gg$ channel becomes more important so that
$R_g^A$ reflects the cross section ratio more closely.  At RHIC, the
PHENIX detector \cite{PHENIX} will use lepton pairs to measure heavy
quark production at $x$ values as low as $3 \times 10^{-3}$.  In this
$x$ region, the $Q^2$ dependence of gluon modifications should be
larger than at the SPS. The nuclear effects vary from 20\% shadowing
for charm to slight antishadowing for bottom.  The increased
importance of the gluon channel reduces the difference between
$d\sigma_{{\rm p}A}/d\sigma_{\rm pp}$ and $R_g^A$ to $\sim 5\%$.

The ALICE detector \cite{ALICE} is designed to study nucleus-nucleus
interactions at the LHC. The high LHC energy makes it ideal for
probing nuclear effects of gluon distributions and studying saturation
effects at low $x$.  For $x$ values as low as $3 \times 10^{-5}$, the
evolution in $Q^2$ should be measurable, as shown in
Fig.~\ref{RGDIST}.  Here, the nuclear modifications are an $\sim 30$\%
effect, and the difference between $d\sigma_{{\rm p}A}/d\sigma_{\rm
pp}$ and $R_g^A$ is of the order of one percent.

All three experiments, taken together, provide a large lever arm in
$\langle x_2 \rangle$ and $\langle Q^2 \rangle$ over which to study
the nuclear gluon distribution.  Measurements at the same energy in
p$A$ and pp interactions with high enough statistics should reveal
much new information on $R_g^A$.  A better determination of $R_g^A$ in
turn provides valuable input to the study of $AA$ interactions.

\bigskip
\noindent {\bf Acknowledgements:} We thank P.V.~Ruuskanen and Z.~Lin
for useful discussions.  R.V. thanks the Niels Bohr Institute for hospitality
during the completion of this work.
The work of K.J.E. and V.K. was supported by the Academy of Finland.
The work of R.V. was supported in part by the Director, Office of
Energy Research, Division of Nuclear Physics of the Office of High
Energy and Nuclear Physics of the U. S.  Department of Energy under
Contract Number DE-AC03-76SF00098.

\end{document}